\documentstyle[preprint,prb,aps]{revtex}
\begin{document}
\draft
\title{Comment on "Critical properties of highly frustrated pyrochlore
         antiferromagnets"}
\author{A. Mailhot and M.L. Plumer}
\address{ Centre de Recherche en Physique du Solide et D\'epartement de
Physique}
\address{Universit\'e de Sherbrooke, Sherbrooke, Qu\'ebec, Canada J1K 2R1}
\date{January 1993}
\maketitle
\begin{abstract}
We argue that the analysis of Reimers {\it et al.} [ Phys. Rev. B {\bf 45},
7295 (1992)] of their Monte Carlo data on the Heisenberg pyrochlore
antiferromagnet, which suggests a new universality class, is not
conclusive.  By re-analysis of their data, we demonstrate
asymptotic volume dependence in some thermodynamic quantities,
which suggests the possibility that the transition may be first order.
\end{abstract}
\pacs{75.40.Mg, 75.40.Cx, 75.10.-b}
%============================================================================
% BODY OF PAPER

Reimers, Greedan and Bj\"orgvinsson \cite{reim} have recently presented
results of Monte Carlo simulations of the pyrochlore Heisenberg
antiferromagnet.  An important issue, as they have emphasised, is to
determine the order of the phase transition.  This highly frustrated
system is interesting from this point of view since Azaria {\it et al.}
\cite{aza} have conjectured that frustration of this type
can easily induce a first-order transition.  Unfortunately, the
determination of the
order of a transition by numerical simulations on finite systems is
not always straightfoward, as emphasized in Ref.[\onlinecite{reim}].
Peczak and Landau,\cite{pec} note that if the correlation length
for a system with a very weak first-order transition
is much larger than the system
size, then pseudocritical behavior may be observed.  Scaling of the
minima near $T_N$ in Binder's
fourth-order energy cumulant \cite{chal},$V=1-\frac13<E^4>/<E^2>^2$,
with system size L (for, e.g. an LxLxL lattice) is believed to be useful for
the characterization of the order of a transition; however, a decisive
conclusion based on this quantity alone can be precarious.  This is
due to the fact that one can never, by simulations of finite systems,
determine if $V^{\ast}_{min}$ is {\it exactly} $\frac23$ or less than
this value, as expected for continuous and first-order transitions,
respectively.  A
value slightly less than $\frac23$ may indicate either a continuous
or very weakly first-order transition.  Nevertheless, it is of
interest to note that the 5-state Potts model in 2D, which is
considered to exhibit one of the weakest first-order transitions known,
has a correlation length estimated to be on the order of
2000 lattice spacings and
meaningful results have been obtained from simulations of
lattices that were a factor of 10 smaller \cite{fuk}.  The extrapolated
value, $V^{\ast}_{min} \simeq 0.6655$, is consistent with expectations.

In addition to $V^{\ast}_{min}$, Reimers {\it et al.} \cite{reim}
considered the scaling behavior of the specific heat and susceptibility
maxima near $T_N$, $C_{max}(L)$ and $\chi_{max}(L)$.  The pyrochlore
lattice contains 16 spins per unit cell and simulations were performed
on lattice sizes L=3-10 (with L=9 omitted).  In terms of more familiar
Bravais lattices with
one spin per unit cell, the number of sites involved correspond
approximately to \~L=$7.6~-~25.2$.  It is certainly not clear
if these lattices
are large enough to reveal the true critical behavior of a highly
frustrated system.  Using data for {\it all} lattice sizes (except
in the case of $V_{min}(L)$ where L=3 results were excluded), they found
exponent ratios estimated to be $\alpha / \nu$=1.8(2) and 1.7(2), from
the scaling of $V_{min}$ and $C_{max}$, respectively, as well as
$\gamma / \nu$=3.5(3) from results for $\chi_{max}$.  In the case
of a first order transition, these thermodynamic quantities should
scale as $L^3$.
They also made the extrapolation $V^{\ast}_{min}$=0.66658(10)
from their data.  With these values,
as well as other scaling results for $\beta$ and $\nu$, they conclude
that the transition is continuous and not within any known
universality class.

Figs. 1, 2 and 3 present an alternative scaling analysis, namely as a
function of $L^3$, of their data
for $\frac23-V_{min}$, $C_{max}$ and $\chi_{max}$, respectively.
The results suggest that the asymptotic behavior of these quantities
are in fact consistent with the transition being first order.  It
appears possible that it
is only for L larger than about 6 or 7 do the simulation results
become meaningful, as indicated by the best-fit straight lines to
data points for these lattice sizes.  The obvious exceptions are the
specific heat and energy cumulant result for L=8, for which we have
no explanation.  Our best-fit estimate for
$V^{\ast}_{min}$ from Fig. 1 is 0.66643(10), noteably further than
from the value $\frac23$ than the result of Reimers {\it et al.}
\cite{reim}.

{}From this analysis, it is clear that the hypothesis of a first
order transition associated with the pyrochlore antiferromagnet
can be taken seriously.  The results of further simulations on
larger lattices would be necessary in order to make a more
definitive statement regarding the critical behavior of this
system.  It is unlikely that experimental work can resolve this
issue.   Even if the transition is in fact weakly first
order, there may well be a significant region of temperature
near $T_N$ where pseudocritical exponents can be measured, with
results consistent with those estimated by Reimers {\it et al.}
\cite{reim}.  This was indeed found to be the case for the exponent
$\beta$.

This work was supported by NSERC of Canada and FCAR du Qu\'ebec.
We thank J. Reimers for useful discussions.

%==============================================================================
% REFERENCES
%

%=============================================================================
%FIGURES
\begin{figure}
\caption{Scaling of the energy-cumulant minima with volume.
The straight line is a fit to the data L=6-10}
\label{fig1}
\end{figure}

\begin{figure}
\caption{Scaling of the specific-heat maxima with volume.
The straight line is a fit to the data at L=6, 7, and 10.}
\label{fig2}
\end{figure}

\begin{figure}
\caption{Scaling of the susceptibility maxima with volume.
The straight line is a fit to the data L=7-10}
\label{fig3}
\end{figure}

%==============================================================================
\end{document}